\documentclass[twoside,twocolumn]{article}
\usepackage{amsmath}
\usepackage{amsfonts}
\usepackage{amssymb}
\usepackage{graphicx}
\usepackage[sc]{mathpazo} 
\usepackage[T1]{fontenc} 
\usepackage{microtype} 

\usepackage[english]{babel} 

\usepackage[hmarginratio=1:1,top=32mm,columnsep=20pt]{geometry} 
\usepackage[hang, small,labelfont=bf,up,textfont=it,up]{caption} 
\usepackage{booktabs} 

\usepackage{lettrine} 

\usepackage{enumitem} 

\usepackage{abstract} 

\usepackage{titlesec} 
\renewcommand\thesection{\Roman{section}} 
\renewcommand\thesubsection{\roman{subsection}} 
\titleformat{\section}[block]{\large\scshape\centering}{\thesection.}{1em}{} 
\titleformat{\subsection}[block]{\large}{\thesubsection.}{1em}{} 

\usepackage{fancyhdr} 
\usepackage{titling} 

\usepackage{hyperref} 

\pretitle{\begin{center}\large\bfseries} 
\posttitle{\end{center}} 
\title{ Deep Recurrent Neural Networks for seizure detection and early seizure detection systems} 
\author{%
\textsc{Sachin Talathi} \\[1ex] 
\normalsize Lawrence Livermore National Lab \\ 
\normalsize \href{mailto:talathi1@llnl.gov}{talathi1@llnl.gov} 
}
\renewcommand{\maketitlehookd}{%
\begin{abstract}
Epilepsy is common neurological diseases, affecting about 0.6-0.8 \% of world population. Epileptic patients suffer from chronic unprovoked seizures, which can result in  broad  spectrum  of  debilitating  medical  and  social  consequences.  
Since seizures, in general, occur infrequently and are unpredictable, automated seizure detection systems are recommended to screen for seizures during long-term electroencephalogram (EEG) recordings. In addition, systems for early seizure detection can lead to the development of new types of intervention systems that are designed to control or shorten the duration of seizure events. In this article, we investigate the utility of recurrent neural networks (RNNs) in designing seizure detection and early seizure detection systems.  We propose a deep learning framework via the use of Gated Recurrent Unit (GRU) RNNs for seizure detection. We use publicly available data in order to evaluate our method and demonstrate very promising evaluation results with overall accuracy close to 100 \%. We also systematically investigate the application of our method for early seizure warning systems. Our method can detect about 98\% of seizure events within the first 5 seconds of the overall epileptic seizure duration.

\end{abstract}
}

\begin{document}
\maketitle


\section{Introduction}

Epilepsy  is a  chronic neurological  disorder characterized  by recurrent, unprovoked seizures \cite{Fisher_2005}. There are 40 to 50 million people with epilepsy worldwide \cite{WHO_2001}.  Currently there is no cure for epilepsy. Many patients' seizures can be controlled,  but  not  cured,  with  medication.  Of those unresponsive to medication, 7\% to 8\% may profit from epilepsy surgery. However, about 25\% of people with epilepsy will continue to experience seizures even with the best available treatment \cite{Mormann:2007aa, Gadhoumi:2016aa}.

The current gold standard for diagnosis of epilepsy is continuous EEG monitoring along with video monitoring of the patient, which usually require in-patient admission. This is an expensive endeavor and may not be always available. In recent years, with the introduction of portable EEG systems, out-patient EEG recordings is becoming quite common. This system has the advantage of EEG being recorded in the patient's natural environment, without any reduction in the seizure frequency, which has been observed in in-patient sessions \cite{Waterhouse:2003aa}. The disadvantage is that screening for seizure from EEG records across multiple days can become onerous task. In these situations, automated seizure detection systems can be extremely useful. 

In addition to seizure detection systems, early seizure warning systems have also become increasingly valuable. There is a growing awareness that controlling seizures might be possible by employing seizure warning based closed-loop treatment strategy \cite{Elger:2001aa}. Early seizure warning systems can also aid patients' to seek safe environment thereby decreasing the risk of injury and the feeling of helplessness that results from seemingly unpredictable seizures. One can also envision an automatic early seizure detection system that can trigger pharmacological intervention in the form of fast-acting drugs \cite{Lin:2017aa} or electrical stimulation \cite{Shenoy:2016aa,Geller:2017aa,Lo:2017aa}. This has the added advantage that the treatment would only occur during an impending seizure. Side effects from treatment with antiepileptic drugs, such as sedation and clouded thinking, could be reduced by on-demand release of a short-acting drug or electrical stimulation during the preictal state.

Research in algorithms for automated seizure detection systems began in early 1970s and over the intervening 45 years, several algorithms have been developed to address this problem \cite{Gotman:1982aa,Gotman:1985aa,Gotman:1999aa,Sirne:1999aa,Fedele:2017aa, Li:2017aa}. Majority of these algorithms work by analyzing the recorded EEG signal(s) to extract relevant information to classify an episode of epileptic seizure from background EEG activity. Given the wide variety of EEG patterns that characterize an epileptic seizure such as `low amplitude desynchronization, polyspike activity, rhythmic waves covering wide range of frequency spectrum and spike waves, and the fact that in extracranial recordings EMG, movement and eye blink artifacts can obscure seizures, several signal-processing techniques have found application in the development of seizure detection systems \cite{Blanco:1996aa,Argoud_2006,Davey:1989aa,Lerner_1996,Schiff:2000aa}. For example, several algorithms have been developed based on spectral or wavelet features \cite{Adeli_2003,Polat_2007}, amplitude measures relative to background \cite{Dingle:1993aa} and chaotic time series measures such as correlation dimension, Lyapunov exponent and entropy \cite{Paivinen:2005aa,Lehnertz:1995aa,Kannathal:2005aa}.

Motivated by recent success of deep learning methods to solve some challenging machine learning problems \cite{Krizhevsky_2012,Mikolov_2011,Hinton_2012}, in this article we investigate the application of recurrent neural network (RNN) model for long temporal sequence learning. Specifically, we design a deep RNN with Gated Recurrent Unit (GRU) hidden units to classify single-channel EEG time series data (each EEG segment being 4097 samples in length, see below for further details) in one of the following three brain-states: 

\begin{itemize}
\item {\bf Healthy} EEG data recorded from a healthy individual
\item {\bf Inter-Ictal} EEG data recorded from an epileptic patient, during non-seizure event
\item {\bf Ictal} EEG data recorded from an epileptic patient during a seizure event
\end{itemize}

\section{Dataset}
The EEG dataset is from the publicly available database on the website of Bonn University \cite{Andrzejak:2001aa}. The dataset includes five subsets, A through E, each containing 100 single-channel EEG segments, each 23.6 s in duration. The EEG segments in A and B are from surface EEG recordings of five healthy volunteers with eyes open and closed respectively. The EEG segments in C an D are from EEG recordings of five epileptic patients, during seizure free intervals (dataset C) and from the hippocampal formation of the opposite brain hemisphere. Finally, EEG segments in S contain seizure activity. EEG segments in C,D and E are from depth electrodes implanted symmetrically into the hippocampal formation. All EEG signals are recorded using the same 128-channel amplifier system and digitized at 173.6 Hz with a 12 bit resolution. Thus, the sample length of each EEG segment is 173.61 x 23.6 $\approx$ 4097 and the corresponding bandwidth is 86.8 Hz.

In Figure 1, we show representative example of raw EEG time-series data from each of the five subsets. In addition, for each example, we plot the auto-correlation function, to demonstrate the time scale of temporal correlations in the data from each of the 3 brain states as well as the spectrogram image for the EEG data. The following observations can be made from the Figure: (a) Data from subsets A and B, corresponding to scalp EEG- recordings from healthy patients have significant line-noise component at 60 Hz, which is absent from depth EEG -recordings from epileptic patients (EEG segments from C,D and E) (b) EEG segment from ictal state (subset E) exhibit activity in the upper-gamma band (40-80 Hz), in contrast to any EEG data from  the other brain-states (c) EEG segment from subset D, corresponding to recordings during seizure free intervals, from the epileptic zone exhibit long-range temporal correlations, which are absent during episodes of ictal events (EEG segment from subset E). 
\begin{figure*}
      \includegraphics[width=1.1\textwidth]{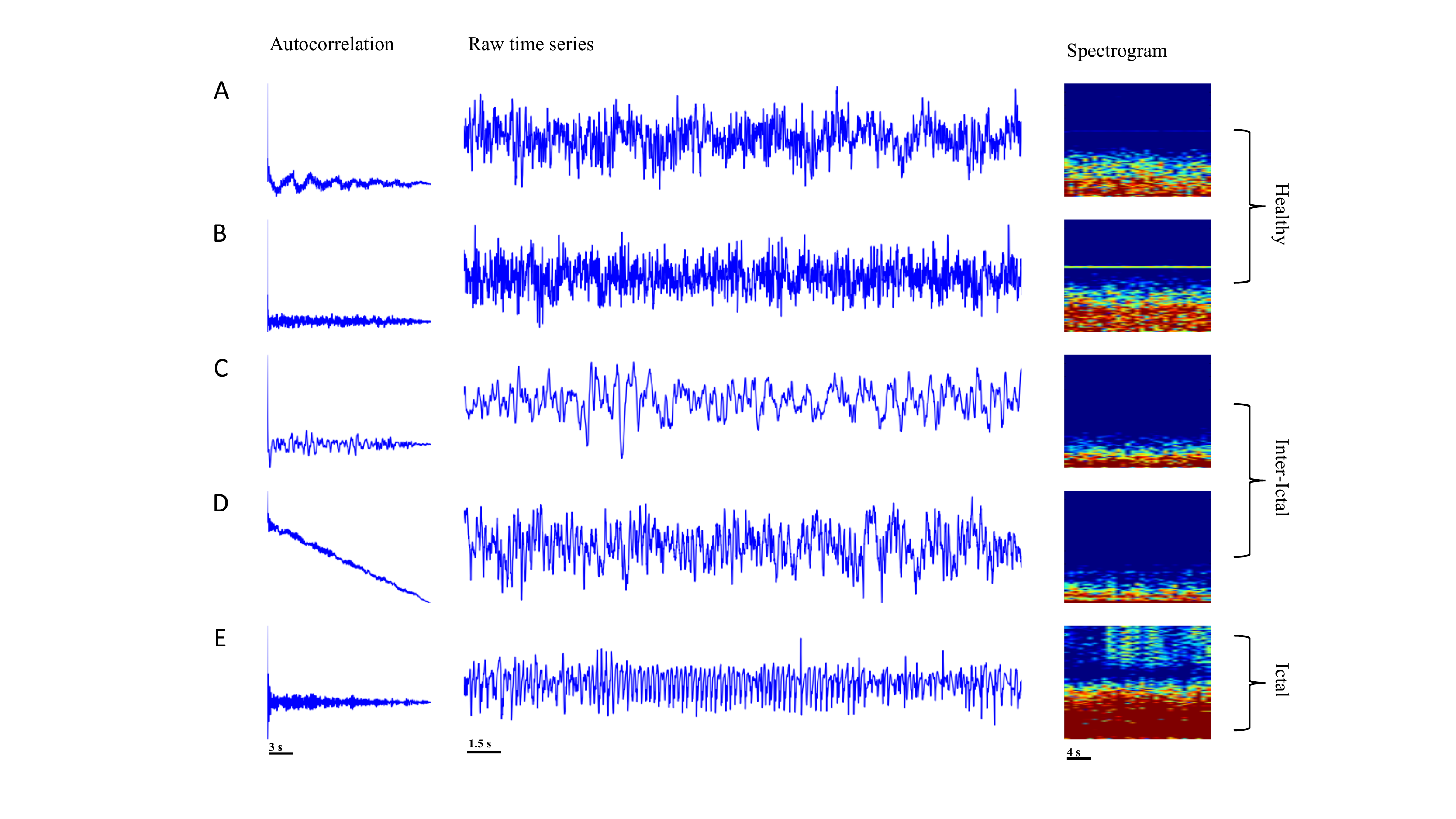}
  \caption{Example of raw EEG segments from the Bonn EEG database. Representative raw-data from each of 5 subsets are plotted and for each example, we plot the spectrogram and the auto-correlation function. }
\end{figure*}

In Figure 2, we plot the distribution of correlation-length (time instance when the auto-correlation function dips below zero for the first time) recorded from the auto-correlation function across all EEG segments available in the dataset. We observe a bimodal distribution, with a significant peak  $\approx$0.5 s and another peak close to 23.6 s. The distribution peak around 23.6 s represents instances where the auto-correlation function never dipped below zero, as is the case for most EEG segments from the subset D.  As noted below, the correlation-length is an important parameter in the design and training of recurrent neural network for EEG-state classification.
\begin{figure}[h]
	\centering
      \includegraphics[width=.5\textwidth]{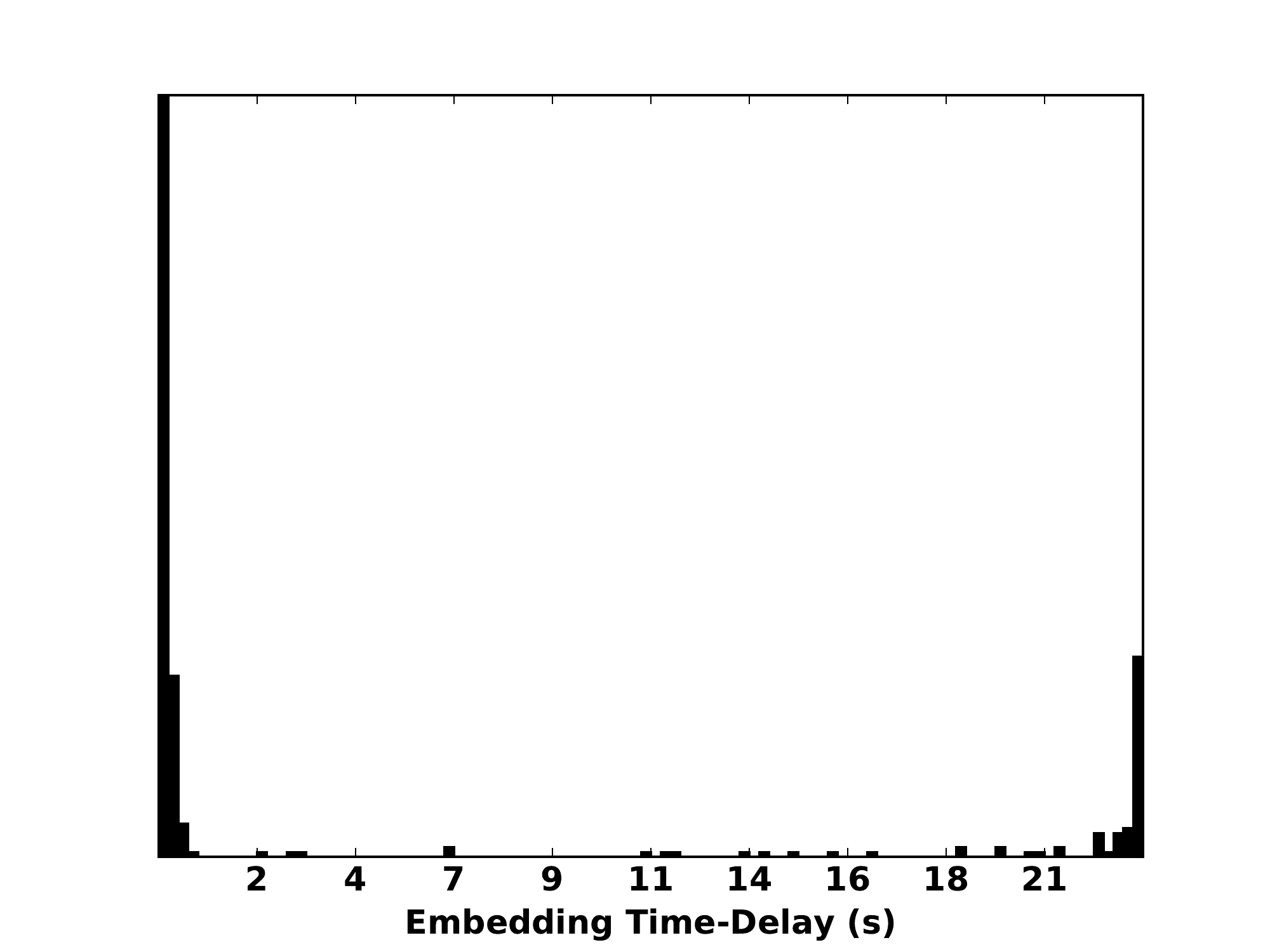}
  \caption{Distribution of embedding time-delays, extracted from auto-correlation function across all EEG segments in the Bonn EEG-dataset}
\end{figure}

\section{Method}
\subsection{Recurrent Neural Networks}
In recent years, Recurrent Neural Networks (RNNs) with sophisticated recurrent hidden units such as the Long-Short-Term Memory (LSTM) unit and the Gated-Recurrent Unit (GRU) have become popular choice for modeling temporal sequences \cite{Graves_2013a, Graves_2013,Fayek:2017aa}. Motivated from these recent successes, here we focus on developing a RNN with GRU-hidden units for solving the EEG classification problem. In what follows, we first provide a brief introduction to RNNs and GRUs. We then describe our proposal for RNN architecture to use for EEG-classification. 

A RNN is a discrete dynamical system with input ${\bf x}_{t}$, an output ${\bf y}_{t}$ and a hidden state ${\bf h}_{t}$.  The dynamical system is defined by 
\begin{eqnarray}
{\bf h}_{t}&=& F({\bf h}_{t-1},{\bf x}_{t}) \nonumber \\
{\bf y}_{t}&=& G({\bf h}_{t}) 
\end{eqnarray}
where $F$ and $G$ are the state transition function and the output function, respectively. 

A conventional RNN is constructed by defining the transition function and the output function as
\begin{eqnarray}
{\bf h}_{t}&=& F({\bf h}_{t-1},{\bf x}_{t}) = \phi_{h}({\bf W}^{T}{\bf h}_{t-1}+{\bf U}^{T}{\bf x}_{t}) \nonumber \\
{\bf y}_{t}&=&G({\bf h}_{t}) = \phi_{o}({\bf V}^{T}{\bf h}_{t})
\end{eqnarray}
where $W$, $U$ and $V$ are the transition, input and output matrices respectively and $\phi_{h}$ and $\phi_{o}$ are
element-wise nonlinear functions. 
Sigmoid or a hyperbolic tangent function are common examples of nonlinear functions used in the construction of convectional RNNs. Back propagation through time (BPTT) is a commonly used stochastic gradient descent algorithm to estimate the parameters of the RNN model. Two  particular  models,  the  long-short term memory (LSTM) RNN \cite{Hochreiter:1997aa} and the  GRU RNN \cite{Cho_2014} have  been  proposed  to  solve  the  ``vanishing'' or  ``exploding''  gradient  problems, which commonly occur in the training of RNNs using BPTT.  Both LSTM-RNN and GRU-RNN use the hidden state from conventional RNN as an intermediate candidate for internal memory cell, say $\tilde{\bf c}_{t}$  and add it in a (element-wise)  weighted-sum  to  the  previous  value  of  the  internal memory state, ${\bf c}_{t-1}$, to produce the current value of the memory cell (state) ${\bf c}_{t} $. The additive memory unit in LSTM and GRU is the key to solving the ``vanishing" or the ``exploding" gradient problem. 
This discrete dynamical equations to represent the LSTM or GRU RNN are given as follows:
\begin{eqnarray}
\tilde{{\bf c}}_{t} &=& \tanh({\bf W}^{T}({\bf r}_{t}\odot{\bf h}_{t-1})+{\bf U}^{T}{\bf x}_{t}) \nonumber \\
{\bf z}_{t}&=&\sigma ({\bf W}^{T}_{z}{\bf h}_{t-1}+{\bf U}^{T}_{z}{\bf x}_{t}+{\bf V}^{T}_{z}{\bf c}_{t-1}) \nonumber \\
{\bf c}_{t} &=& {\bf f}_{t}\odot {\bf c}_{t-1}+ {\bf i}_{t}\cdot \tilde{\bf{c}}_{t} \nonumber \\
{\bf h}_{t}&=&{\bf o}_{t}\odot \phi_{o}({\bf c}_{t}) 
 \end{eqnarray}
 where ${\bf z}=\{{\bf i},{\bf f},{\bf o},{\bf r}\}$, representing the gating functions: input gate, the forget gate, the output gate and the internal gate and $\sigma$ is the Sigmoid function.  The trainable model parameters are: $\{{\bf W}, {\bf W}_{z}, {\bf U}, {\bf U}_{z}, {\bf V}_{z}\}$
 
 For LSTM unit, typical choice is ${\bf r}_{t}=\mathbb{I}$, whereas for GRU unit we set, ${\bf V}_{z}=0$, ${\bf h}_{t}={\bf c}_{t}$ and ${\bf f}_{t}=1-{\bf i}_{t}$.  
 
 \subsection{Proposed Classification Method}
 
 We specifically focus on the 3-class classification problem of classifying a given EEG segment into one of the healthy, inter-ictal or ictal states. Accordingly, we collate the EEG data from 5 subsets as follows: $\text{Healthy}=\{\text{A},\text{B}\}$; $\text{Inter-Ictal}=\{\text{C},\text{D}\}$ and $\text{Ictal}=\text{E}$. As there are 100 EEG segments for each of the subsets A,B,C, D and E, half of the segments from each subset, randomly chosen, are used for training and the remainder half are used for testing. Thus, both the training and the testing data is comprised of a total of 250 EEG segments, 100 each for the healthy and inter-ictal states and 50 for the ictal state.   
 
Each EEG segment in the dataset is comprised of 4097 data samples ($23.6 \times 173.61$). Training of RNN on such a long-temporal sequence is quite challenging. In order to facilitate RNN training, we exploit the correlations in the data (see Fig 2) and divide the long temporal sequence of 4097 sample EEG segment into 51 sub-segments, each 80 samples long. We discard the last 17 samples from each EEG data record. Each sub-segment is about 0.46 s in duration, corresponding to the dominant peak in the correlation-length of the dataset. For each EEG sub-segment, we assign the same label, derived from the label for the underlying EEG segment. This procedure of pre-processing the dataset to make it amenable for training with RNNs is demonstrated in a schematic diagram in Figure 3. 

\begin{figure}[h]
	\centering
      \includegraphics[width=.5\textwidth]{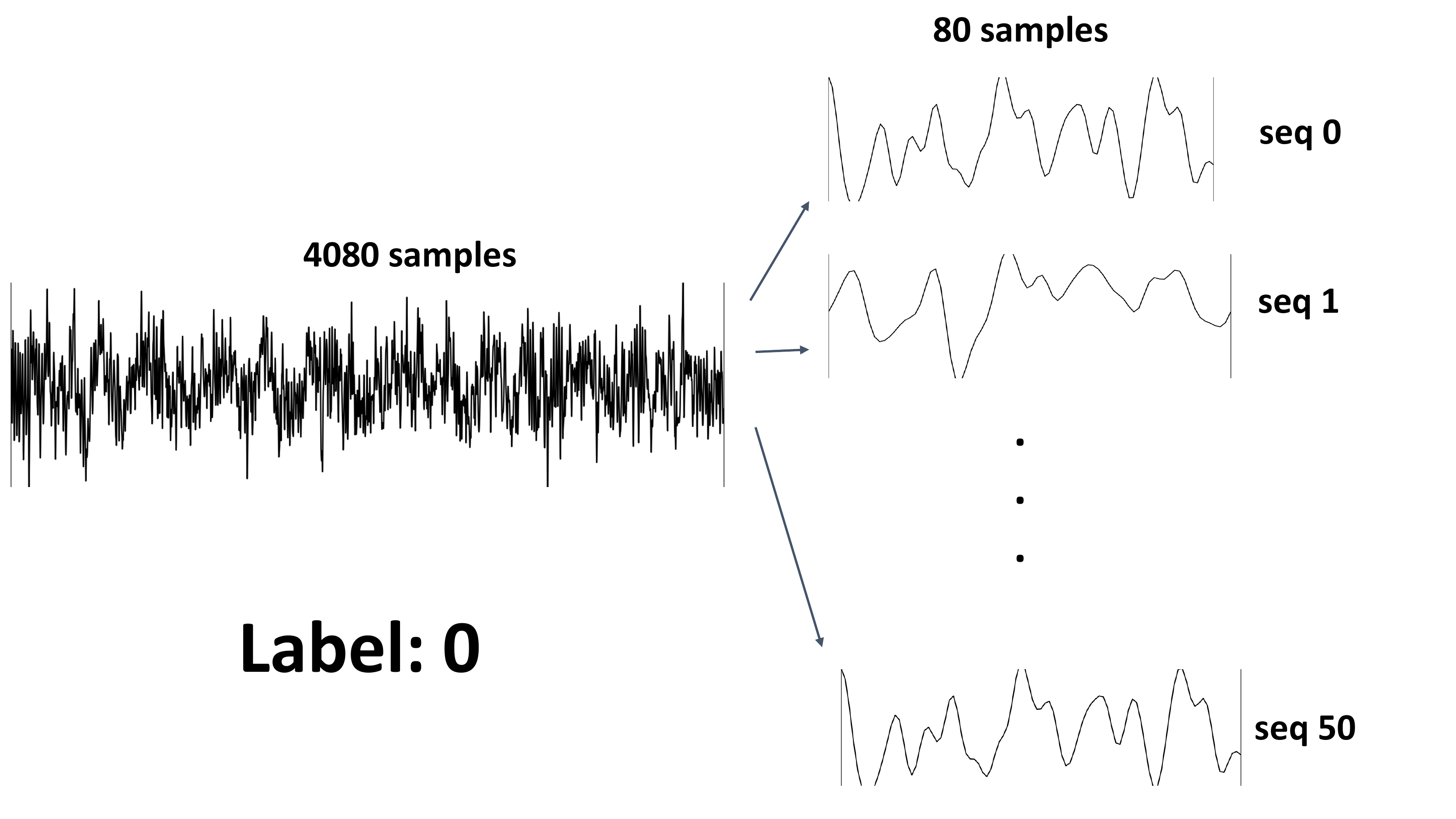}
  \caption{Schematic diagram demonstrating how the original EEG segment (4080 samples) is split into equal length sub-segments (80 samples per sub-segment ), with each sub-segment sharing the same label of the original EEG segment}
\end{figure}

The specific RNN model that we adopt for training is schematically described in Table \ref{Table1}. The model is comprised of a hidden recurrent layer with 100 GRU units, which is followed by a fully connected (fc) layer with 100 hidden nodes. We use linear activation function for the fc layer. The fc layer is applied in a time-distributed fashion across the entire length of the input sequence. The output of the fc layer is fed into another recurrent layer with 100 GRU units. The output of this second hidden recurrent layer is fed into the logistic regression layer to predict the class probability for the input data sequence. The RNN model prediction for presence or absence of a seizure event in a given EEG segment  is estimated by averaging the model predictions across all 51 EEG sub-segments. 

\begin{table}[t]
\label{my-label}
\begin{tabular}{|l|l|l|}
\cline{1-3}
Layer Type                       & Output Shape & Parameters \\\cline{1-3}
Input                            & (51, 80, 1)  & 0                 \\\cline{1-3}
GRU 0           & (51,100, 1)  & 30600             \\\cline{1-3}
Fc & (51,100,1)   & 10100             \\\cline{1-3}
GRU 1           & (51,100)     & 60300             \\ \cline{1-3}
LR              & (51,3)       & 303               \\ \cline{1-3}
\end{tabular}
\caption{Recurrent Neural Network Model Architecture. GRU: Gated Recurrent Unit hidden layer; Fc: Fully Connected hidden layer; LR: Logistic Regression classification layer with softmax non-linearity}
\label{Table1}
\end{table}

\subsection{RNN model training}
We use the default initialization parameters from keras package for initializing the  weights of GRU hidden units as well as those of the fully connected layer in the RNN  model. We train the RNN in stateful-mode, implying that information from previous state of the internal memory unit is propagated across samples in the batch containing the training sequence.   

 The internal memory state for each of the GRU unit is initialized to 0. While we divide the original EEG segment of 4097 data samples into sub-segments, at the training time, for each sub-segment, the state of the memory cell resulting from weight updates from previous sub-segment is preserved.  All training is performed using Adam stochastic optimization \cite{Kingma_2014} and employ clipping of the gradients \cite{Pascanu_2014}. For all model training we begin with learning rate of 0.01 and perform model training for 300 epochs, rescaling the learning rate by factor 0.1 at each 100th epoch. 

\section{Results}
\begin{figure}[t]
\includegraphics[width=.6\textwidth]{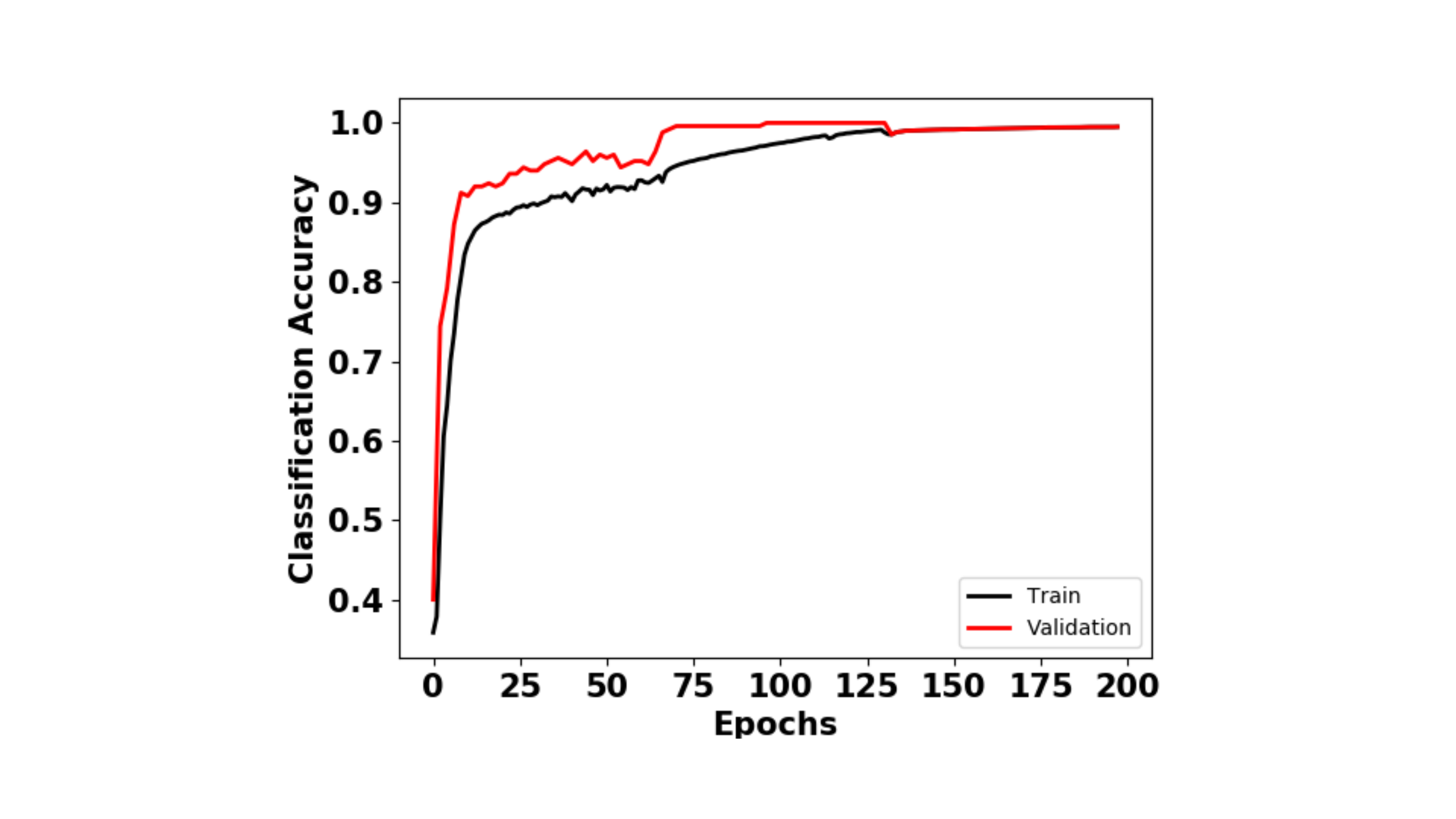}
  \caption{Plot of training (and validation) accuracy for GRU-RNN as function of training epochs}
\label{Fig4}
\end{figure}

 In Figure \ref{Fig4}, we plot the seizure-detection classification accuracy for the RNN model as a function of training epochs. We note that the model achieves accuracy of 99.6 \% on the validation dataset.  

 \begin{table}[]

\begin{tabular}{|l|l|l|l|}
\cline{1-4}
Class                & \textbf{Healthy} & \textbf{Inter-Ictal} & \textbf{Ictal} \\ \cline{1-4} 
\textbf{Healthy}     & 100              & 0                    & 0              \\ \cline{1-4} 
\textbf{Inter-Ictal} & 1                & 99                   & 0              \\ \cline{1-4} 
\textbf{Ictal}       & 0                & 0                    & 50             \\ \cline{1-4} 
\end{tabular}
\caption{Confusion-matrix for model prediction on validation EEG dataset}
\label{Table2}
\end{table}

\begin{figure*}[t]
 \centering
\includegraphics[width=.75\textwidth]{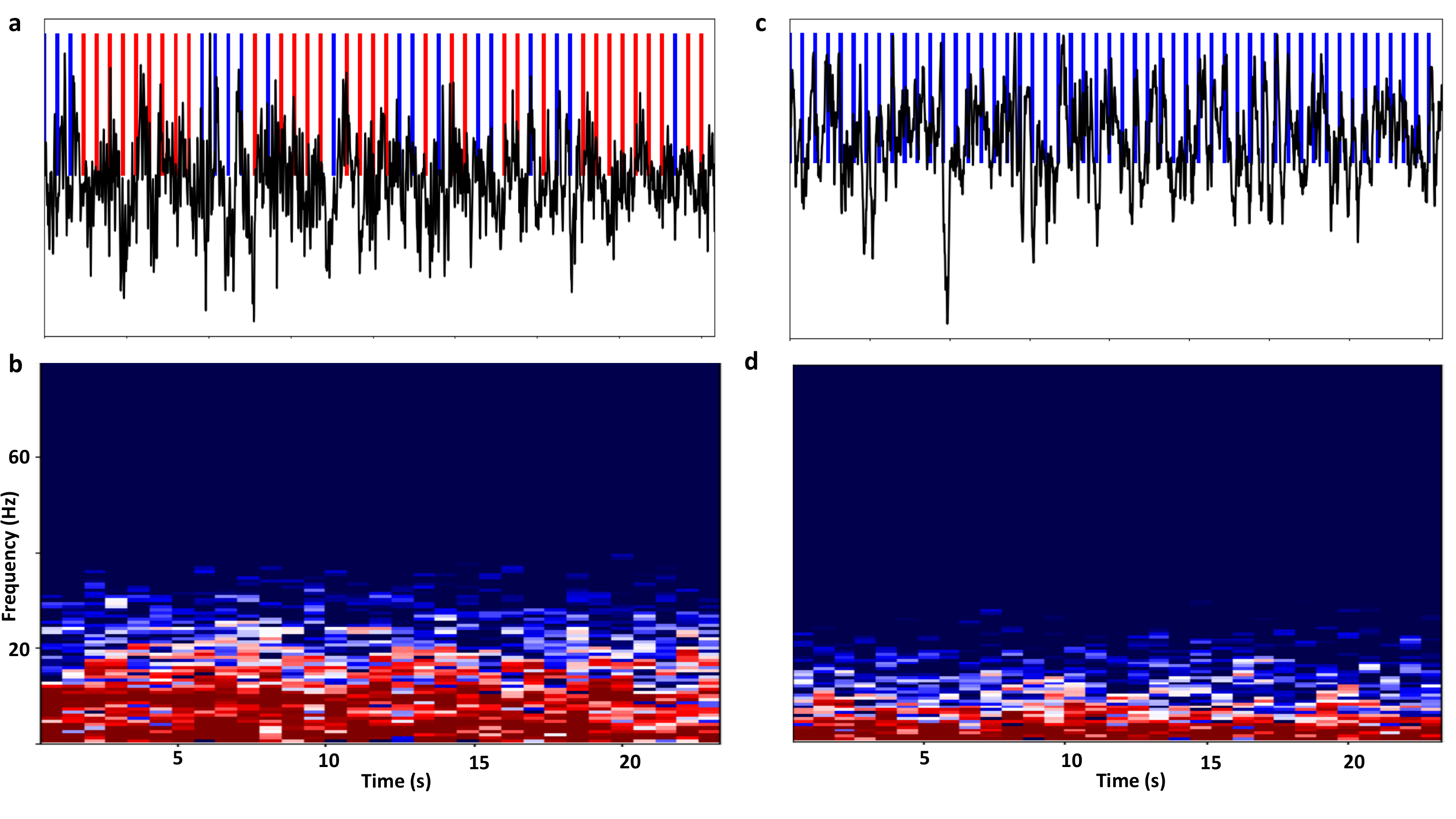}
  \caption{EEG time-series trace and the corresponding spectrogram of an inter-ictal EEG segment predicted incorrectly by the GRU-RNN model (in a and b) and predicted correctly by the GRU-RNN model (in c and d).  The EEG sub-segments that were correctly predicted by the GRU-RNN model are marked by blue vertical bars, whereas the EEG sub-segments that were incorrectly predicted as belonging the healthy EEG segment are marker by red vertical bars. It is clear from visual inspection of the respective spectrograms (b and d) that, the spectral features of inter-ictal EEG segment that was predicted incorrectly has features quite similar to those of healthy EEG segment rather than those belonging to inter-ictal EEG segment.}
\label{Fig5}
\end{figure*}

 The confusion matrix for model predictions on validation data set is presented in Table \ref{Table2}.
 The model is able to predict 100 \% of healthy and epileptic EEG segments and produced 1 error on the inter-ictal EEG segment. We probe the wrongly predicted EEG-segment in Figure \ref{Fig5}.

 In Figure \ref{Fig5}a, we show the raw time trace of the wrongly predicted EEG segment and overlay on top the predictions for the sub-segments. We notice that majority of sub-segments are predicted to belong to healthy EEG, resulting in overall prediction for the EEG segment to belong to healthy EEG class. In Figure \ref{Fig5}b, we plot the spectrogram for the wrongly predicted EEG segment. Qualitatively, the spectrogram looks much similar to that of a healthy EEG segment.  For comparison, in Figure \ref{Fig5}c, we also show an example of correctly predicted inter-ictal EEG segment with overlay of model predictions for each sub-segment of the chosen EEG segment and Figure \ref{Fig5}d, we show the corresponding spectrogram. From the spectrograms in \ref{Fig5}b and \ref{Fig5}d, it is clear that the model captures the low frequency dominant signal features, characteristics of the inter-ictal EEG segment, where the single error stems from the fact that the particular signal has spectrum characteristics more similar to those of healthy EEG segment. 
 
\begin{figure}[t]
\includegraphics[width=.6\textwidth]{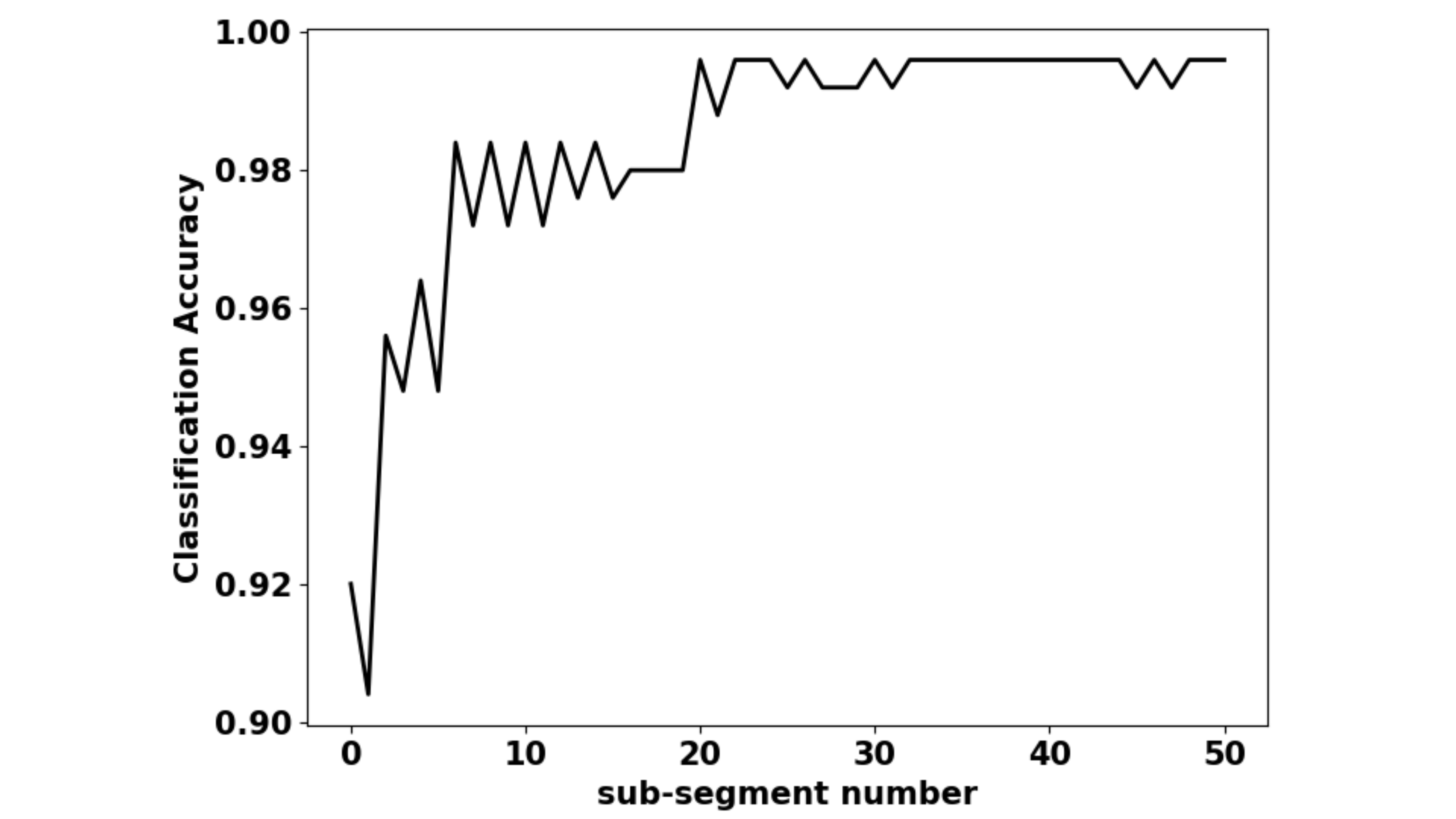}
  \caption{Plot of GRU-RNN model accuracy on validation dataset as function of the number of EEG sub-segments used to predict the class label for the EEG segment.}
\label{Fig6}
\end{figure}

For early seizure detection, the time into a seizure event when the algorithm triggers a seizure event is very important. We quantify the performance of GRU-RNN model for early detection by plotting the model accuracy as function of the number of EEG sub-segments used to make a decision on the category to which the EEG segments belongs. The result is depicted in Figure \ref{Fig6}. We see that GRU-RNN model is able to correctly predict the EEG segment class labels with $\approx$ 98 \% by using about 10 EEG sub-segments, each of length $\approx$ 0.46 s in duration. In other words, within 5 seconds of the initiation of seizure event, the GRU-RNN model is able to predict the seizure event with about 98\% accuracy. 

This findings offers a strong support to the utility of GRU-RNN model for use in early-seizure detection system that can be extremely useful for developing closed loop seizure control systems where timely intervention can be leveraged to abate seizure progression.

\section{Conclusion}
In this paper, we explored the ability for a deep RNN model to classify EEG segments which contain epileptic seizures. We present a novel GRU-RNN model, that can be trained with very high degree of accuracy to classify EEG segments belonging to one of the three: healthy, inter-ictal and ictal states. Previous published state-of-the-art results for the 3-class classification of EEG segments on the Bonn EEG dataset offered accuracy of about 98\% \cite{Tzallas_2007}. Our proposed method offers a new state-of-the-art classification performance of close to 100\% accuracy for this task of EEG state classification. Furthermore, we present results to demonstrate the utility of our proposed method in developing early seizure-detection systems. 

Our proposed GRU-RNN seizure detection system offers several advantages over existing algorithms for seizure detection. For example, we work directly with the raw input data that is minimally pre-processed. 
The model is light weight, with on the order of 100,000 trainable parameters. It remains to be seen how our model will scale when applied to much larger multivariate EEG datasets. 

\bibliographystyle{ieeetr}
\bibliography{References} 
\end{document}